\documentclass[useAMS,usenatbib]{mn2e}
\usepackage{latexsym,graphicx,natbib}
\usepackage{amsmath}

\newcommand\msun{\, \rm M_\odot}
\newcommand\pc{{\, \rm pc}}

\newcommand{\gap}{\;\rlap{\lower 2.5pt \hbox{$\sim$}}\raise 1.5pt\hbox{$>$}\;}
\newcommand{\lap}{\;\rlap{\lower 2.5pt \hbox{$\sim$}}\raise 1.5pt\hbox{$<$}\;}

\newcommand\thard{\, \rm T_h}

\title[Massive black hole binary plane reorientation]
{Massive black hole binary plane reorientation in rotating stellar systems}

\author[Gualandris et al.]  
  {Alessia Gualandris$^{1}$\thanks{E-mail: alessiag@mpa-garching.mpg.de}, Massimo Dotti$^{2}$, and Alberto
  Sesana$^{3}$\\ 
  $^{1}$Max-Planck Institut f\"{u}r Astrophysik, Karl-Schwarzschild-Str. 1, D-85741 Garching, Germany\\
  $^{2}$Dipartimento di Fisica G.~Occhialini, Universit\`a degli Studi
  di Milano Bicocca, Piazza della Scienza 3, 20126 Milano, Italy\\
  $^{3}$Albert Einstein Institut, Am M\"{u}hlenberg 1, Golm, D-14476,
  Germany.\\
 }

\begin{document}

\date{}

\maketitle

\begin{abstract}
  We study the evolution of the orientation of the orbital plane of
  massive black hole binaries (BHBs) in rotating stellar systems in
  which the total angular momentum of the stellar cusp is misaligned
  with respect to that of the binary. We compare results from direct
  summation $N$-body simulations with predictions from a simple
  theoretical model. We find that the same encounters between cusp
  stars and the BHB that are responsible for the hardening and
  eccentricity evolution of the binary, lead to a reorientation of the
  binary orbital plane. In particular, binaries whose angular momentum
  is initially misaligned with respect to that of the stellar cusp
  tend to realign their orbital planes with the angular momentum of
  the cusp on a timescale of a few hardening times. This is due to
  angular momentum exchange between stars and the BHB during close
  encounters, and may have important implications for the relative
  orientation of host galaxies and radio jets.

\end{abstract}

\begin{keywords}
black hole physics -- methods: numerical -- stellar dynamics
\end{keywords}

\section{Introduction}

In the standard $\Lambda$CDM cosmological scenario, massive black hole
binaries (BHBs) are the natural outcome of mergers of massive galaxies
\citep{BBR80}. The orbital decay of BHBs has been studied in detail
under the assumption of purely stellar systems \citep[e.g.][]{mf2004,
  ber05, baum06, mms07, sesa2007, sesa2010} as well as gaseous
environments \citep[e.g.][]{armitage02, armitage05, escala04, escala05,
  dotti06, dotti07, dotti09, cuadra09, lodato09, roedig11}.

Most studies of the evolution of BHBs in stellar environments
consider non rotating, spherically symmetric nuclei, despite the
natural occurrence of rotation in merger remnants
\citep[e.g.][]{gm11}.  Simulations of MBHs in spherical stellar
systems predict a stalling of the binary evolution at parsec scale
separations, the ``final parsec problem'' \citep{mm03}, which may or
may not be avoided in the presence of gas dissipation.  A BHB can,
however, reach coalescence even in a completely gas poor environment,
if embedded in a non-spherically symmetric stellar distribution.
Purely dissipationless simulations of non-spherical galaxy models
\citep{ber06,ber09} or galaxy mergers \citep{khan11, preto11, gm11}
find continuing hardening of the binary down to separations where
energy loss due to emission of gravity waves becomes
efficient. Simulations by \citet{gm11} show that efficient hardening
is achieved even in systems close to axisymmetry. This departure from
spherical symmetry is due to the rotation introduced by the merger,
which results in an enhanced rate of stars interacting with the
binary.

A comprehensive and systematic study of the evolution of BHBs in
rotating stellar cusps is still missing. \citet{sgd11} (hereafter
paper I) studied the evolution of the eccentricity of unequal mass
BHBs in rotating systems by means of an hybrid analytical/3-body
scattering formalism as well as full $N$-body simulations. They found
a strong dependence of the eccentricity evolution of the binary on the
degree of co-rotation in the stellar cusp, with binaries increasing
their eccentricity in cusps containing a significant fraction of
counter-rotating stars.  This is relevant to the case of galactic
minor mergers, in which counter-rotating systems can be produced as a
result of rotation in the larger galaxy, thereby forming very
eccentric binaries.

Another important and potentially observable parameter is the
orientation of the binary orbital plane. Its evolution has been
studied in spherical and isotropic systems \citep{m02, gm07}. Due to
the lack of any preferential direction in these systems, the orbital
plane of the binary can only undergo a random walk, resulting in small
changes in the orbital plane on long timescales. Here we study the
evolution of the orbital plane of BHBs in rotating stellar systems,
where, as will be discussed, a significantly larger reorientation is
expected.

The letter is organized as follows. In section\,\ref{sec:spherical} we
describe the evolution of the binary orbital plane in spherically
symmetric models. In section\,\ref{sec:rotating} we describe the case
of rotating models, comparing results from $N$-body simulations with a
simple analytical model.  Final remarks are presented in
Section\,\ref{sec:concl}.

\section{Non rotating cusps}
\label{sec:spherical}

The orientation of the orbital plane of BHBs embedded in non
rotating, isotropic stellar nuclei has been studied for the first time
by \citet{m02}. In such systems, repeated encounters with passing
stars cause the orientation of the binary to undergo a random walk (a
``rotational Brownian motion'', in \citet{m02}). From simple analytical
arguments, the expected change in orientation on one hardening time
$\thard = \left|a / \dot{a}\right|$ of the binary (which is of the order
of $120$ initial binary orbital periods) is  \citep{gm07}
\begin{equation}
\label{eq:deltat} \Delta \theta \propto q^{-1/2}
\left(\frac{m_*}{M_{\rm BHB}}\right)^{1/2} \left(1-e^2\right)^{-1/2}
\end{equation}
where $q = M_2/M_1$ is the binary mass ratio, $M_{\rm BHB}$ the total
mass, and $e$ the orbital eccentricity. Numerical scattering
experiments have constrained the normalization in Eq.~\ref{eq:deltat}
only for circular binaries. Neither the dependence on the eccentricity
nor on the mass ratio has been tested numerically. In the case of a
binary with $M_{\rm BHB} = 10^6 m_*$ and $q=10^{-3}$, the expected
reorientation is $\Delta \theta \approx 9^{\circ}$.

Here we check the validity of Eq.~\ref{eq:deltat} by performing
$N$-body simulations of binaries with different eccentricities (in the
range $0-0.99$) embedded in non-rotating stellar cusps. We use the
direct summation $N$-body code $\phi$GRAPE \citep{Harfst2007} in
combination with the $\tt Sapporo$ library \citep{sapporo2009} to
accelerate the computations on GPU hardware.  In our integrations we
adopted $M_1 = 10^6\msun$ and $q=1/81$ and set the initial semi-major
axis to $a_i = 0.06\pc$. The MBHs are embedded into a stellar cusp
following a Bahcall-Wolf $\rho(r) \sim r^{-7/4}$ density profile at
distances smaller than $1\pc$, with total mass $M_{c} \sim 2.5\times
10^5 \msun$ and a mass enclosed in the binary orbit equal to $2
M_2$. The cusp is modelled with $N=32\rm k$ equal-mass particles,
resulting in a black hole to star mass ratio of $ m_* / M_1 =
7.5\times10^{-6}$.
\begin{figure}
  \begin{center}
    \includegraphics[width=8cm]{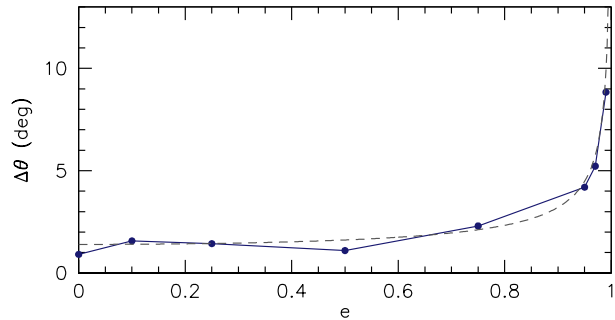}
  \end{center}
  \caption{Change over one hardening time in the orientation of the
    angular momentum of the binary, as a function of the initial
    eccentricity of the BHB. The dashed line represents the
    theoretical dependence given in Eq.~\ref{eq:deltat}, normalized to
    match the data at $e=0.25$. The amplitude of the reorientation in
    our runs is however comparable to the analytical estimates in
    \citet{m02}.}
  \label{fig:ecc}
\end{figure}
Fig~\ref{fig:ecc} shows the degree of re-alignment of the binary plane
as a function of the initial eccentricity. The dashed line indicates
the dependence expected from the analytical model, while the points
show the results of our $N$-body integrations. The agreement is
remarkable.

For a given binary, the re-orientation in a non-rotating system does
not depend only on the mass density and velocity field of the stars,
but also on the mass ratio between the binary and a single star in the
cusp\footnote{Note that this ratio is properly defined only for a
  single mass stellar cusp. In the presence of a mass spectrum, the
  mean mass of the interacting stars is a good proxy for $m_*$.}.  A
binary embedded in a stellar cusp would, on average, experience a
greater realignment as a result of few encounters with massive stars
than as a result of many encounters with low-mass stars. This feature
is typical of any random walk process, in which there is no
preferential direction for the change of the binary plane in a single
encounter and changes due to different encounters tend to cancel each
other out. The same random walk behaviour is observable in rotating
cusps, if the angular momentum of the binary is aligned with that of
the cusp \citep{pau2010}.  This behaviour is illustrated in the lower
panel of Fig~\ref{fig:Ndep}, which shows the dependence of the change
in $\theta$ on the number of particles used to model the cusp, i.e.,
on the $m_*/M_{\rm BHB}$ mass ratio. The agreement between the
numerical results (full circles) and Eq.~\ref{eq:deltat} (solid line)
is outstanding.

\section{Rotating cusps}
\label{sec:rotating}

\begin{figure}
  \begin{center}
    \includegraphics[width=8cm]{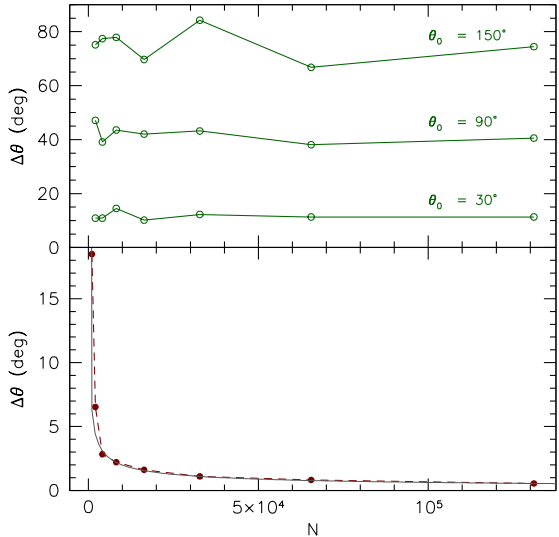}
  \end{center}
  \caption{Change in the binary plane orientation per hardening time
    versus number of particles in the cusp (in the range $N=2 \rm k$
    to $N=128 \rm k$) or, equivalently, ratio
    between star mass and MBH mass. In all the runs $e=0.5$. The lower
    panel refers to
    isotropic models, while
    the upper panel refers to rotating models with ${\cal F} = 0.875$
    and an initial angle between the binary angular momentum and that
    of the stellar cusp $\theta_0 = 30^{\circ}, 90^{\circ},
    150^{\circ}$.}
  \label{fig:Ndep}
\end{figure}

Here we study in detail the re-orientation of a massive BHB embedded
in a cusp with net rotation. To this aim, we perform high accuracy
$N$-body simulations similar to those presented in the previous
section, the only difference being the degree of rotation that we
enforce in the cusp. This is obtained, as in Paper I, by reversing the
sign of all velocity components for a random subset of cusp stars, at
the time where the BHB is added.  In particular, we generated models
with a fraction ${\cal F} = 0.875$ of co-rotating stars and varied the
initial angle $\theta_0$ between the total angular momentum of the
cusp and the angular momentum of the binary. We considered cases with
$\theta_0 = 0^{\circ}, 30^{\circ}, 60^{\circ}, 90^{\circ},
120^{\circ}, 150^{\circ}.$

\begin{figure}
  \begin{center}
    \includegraphics[width=8cm]{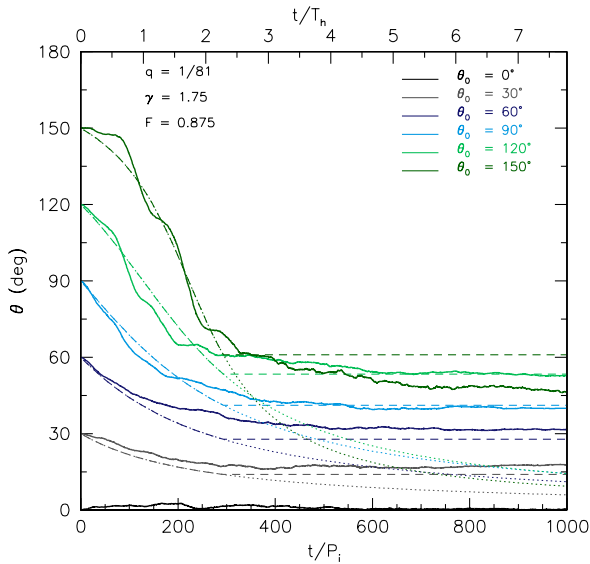}
  \end{center}
  \caption{Evolution of the angle $\theta$ between the angular
    momentum vector of the binary and that of the stellar cusp.  Time
    is expressed in units of the initial binary orbital period
    (bottom) and of the hardening time (top). Different solid lines
    are for models with different initial values of $\theta$, from
    $\theta_0=0^{\circ}$ to $\theta_0=150^{\circ}$. Dotted lines
    represent the predictions from the theoretical model while dashed
    lines indicate the predictions modified to take into account the
    finite reservoir of angular momentum in the stellar cusp.}
  \label{fig:theta}
\end{figure}
The evolution of $\theta$ as a function of time is shown in
Fig.~\ref{fig:theta} (solid lines) for all runs. Binaries whose
angular momentum is initially mis-aligned with respect to that of the
cusp tend to realign it on a time-scale of a few hardening times.  The
re-alignment can be quite significant, with an average change in
$\theta$ of the order of 50\% of the initial value, and up to 70\% in
some cases. For example, the case with $\theta_0 = 150^{\circ}$
reaches a final value of $\theta \sim 45^{\circ}$, with a change of
more than $100^{\circ}$ over one hardening time. This is about 100
times larger than the reorientation measured for a similar isotropic
model (see Fig.~\ref{fig:ecc}). As will be discussed below, such large
reorientation of the orbital plane may have observational
consequences. However, we note that the re-alignment is not complete
but stops at a finite value of $\theta$.

\begin{figure}
  \begin{center}
    \includegraphics[width=8cm]{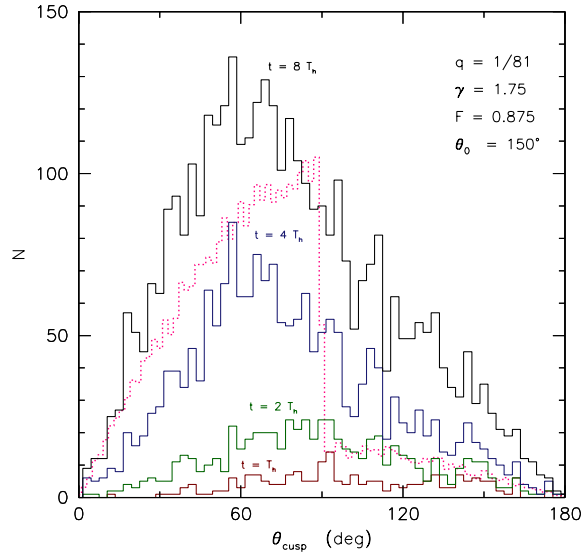}
  \end{center}
  \caption{Solid lines: Distribution of the angle $\theta_{\rm cusp}$
    between the angular momentum vector of a single star and the
    initial angular momentum vector of the cusp, for the run with
    $\theta_0= 150^{\circ}$. Only unbound stars are considered. The
    different curves refer to different times during the evolution,
    from $1\thard$ to $8\thard$. Stars are ejected roughly
    isotropically from early times. Dotted line: Initial distribution
    of $\theta_{\rm cusp}$ for all cusp stars, normalized by a factor
    of 10, for an easy comparison.}
  \label{fig:thetacusp}
\end{figure}
The evolution of $\theta$ can be explained in terms of a simple
analytical model. Let $\bmath{l}_B = \bmath{L}_B / \mu$ be the angular
momentum per unit mass of the binary, with $\mu=M_1 M_2 /(M_1+M_2)$
the reduced mass, and $\bmath{l}_{\star} = \bmath{L}_{\star} /
m_{\star}$ the angular momentum per unit mass of a star interacting
with the binary. Conservation of angular momentum during an encounter
(assuming $m_{\star} \ll M_1+M_2$) gives
\begin{equation}
\label{deltalb}
\Delta \bmath{l}_B = \frac{m_{\star}}{\mu}  \Delta \bmath{l}_{\star} \,.
\end{equation}

Considering that, once normalized per unit mass, the average change
per encounter in the stellar angular momentum is of the order of the
binary angular momentum \citep{m02}\footnote{This allows us to ignore
  the evolution of the magnitude of the binary angular momentum due to
  hardening when modelling the evolution of the plane orientation. The
  change in $\theta$ would be the same even if we considered the decay
  of the binary orbit.}, each interaction contributes, on average, a factor
\begin{equation} 
< \Delta \bmath{l}_B > \sim \frac{m_{\star}}{\mu} \bmath{l}_B
\end{equation} 
to the binary angular momentum. The rate at which $\bmath{l}_B$ varies with
time is
\begin{equation} 
\frac{{\rm d} \bmath{l}_B}{{\rm d} t}= \frac{{\rm d} n_{\rm enc}}{{\rm
    d} t} < \Delta \bmath{l}_B > \,, 
\end{equation}
where ${\rm d} n_{\rm enc}/{\rm d} t$ is the rate of binary/star
interactions. 
The rate can be estimated as
\begin{eqnarray}
\label{dndt}
\frac{{\rm d} n_{\rm enc}}{{\rm d} t} & = & \pi ~ n_* R_{\rm i,2}^2 ~ v_{\rm rel}\nonumber \\
 & \approx &  \pi ~ n_* \left(\frac{G M_2}{v_{\rm rel}^2}\right)^2  v_{\rm rel}\nonumber \\
 & \approx &  \pi ~ n_* \left(\frac{M_2}{M_1}\right)^2  \sqrt{G M_1 a^3}\,,
\end{eqnarray}
where $n_*$ is the number density of stars in the cusp, $R_{\rm i,2}$
is the influence radius of the secondary hole, and $v_{\rm rel}$ is
the mean relative velocity between the stars and $M_2$.  Note that we
have implicitly assumed that: {\it i) only stars efficiently
  interacting with the secondary modify the angular momentum of the
  binary; ii) the dynamics outside the influence radius of the
  secondary is completely dominated by the primary}.

In order to apply Eqs.\,\ref{deltalb}-\ref{dndt}, we need to specify
the average direction of $<\Delta \bmath{l}_{\rm B}>$.  This is
opposite to the direction of the variation of the angular momentum of
the star during the interaction.  In a reference frame in which the
$z$ direction is aligned with the total angular momentum of the cusp,
the average $x$ and $y$ components of the angular momentum of the
stars before an encounter are zero.  In this simple model we assume
that {\it the final angular momentum of a star that experienced an
  interaction is isotropically distributed}.  This is justified by the
observation that, at any time, stars after an interaction with the
binary are much more isotropically distributed than before, as shown
in Fig.~\ref{fig:thetacusp} for the population of unbound stars.  As a
consequence, the variation $<\Delta \bmath{l}_* > =
<\bmath{l}_{*,1}-\bmath{l}_{*,0}>$ points in the direction opposite to
the angular momentum of the cusp $\bmath{l}_{\rm cusp}$, and $<\Delta
\bmath{l}_B>$, averaged over repeated encounters, points in the $z$
direction, parallel to $\bmath{l}_{\rm cusp}$.
\begin{figure}
  \begin{center}
    \includegraphics[angle=90,width=8cm]{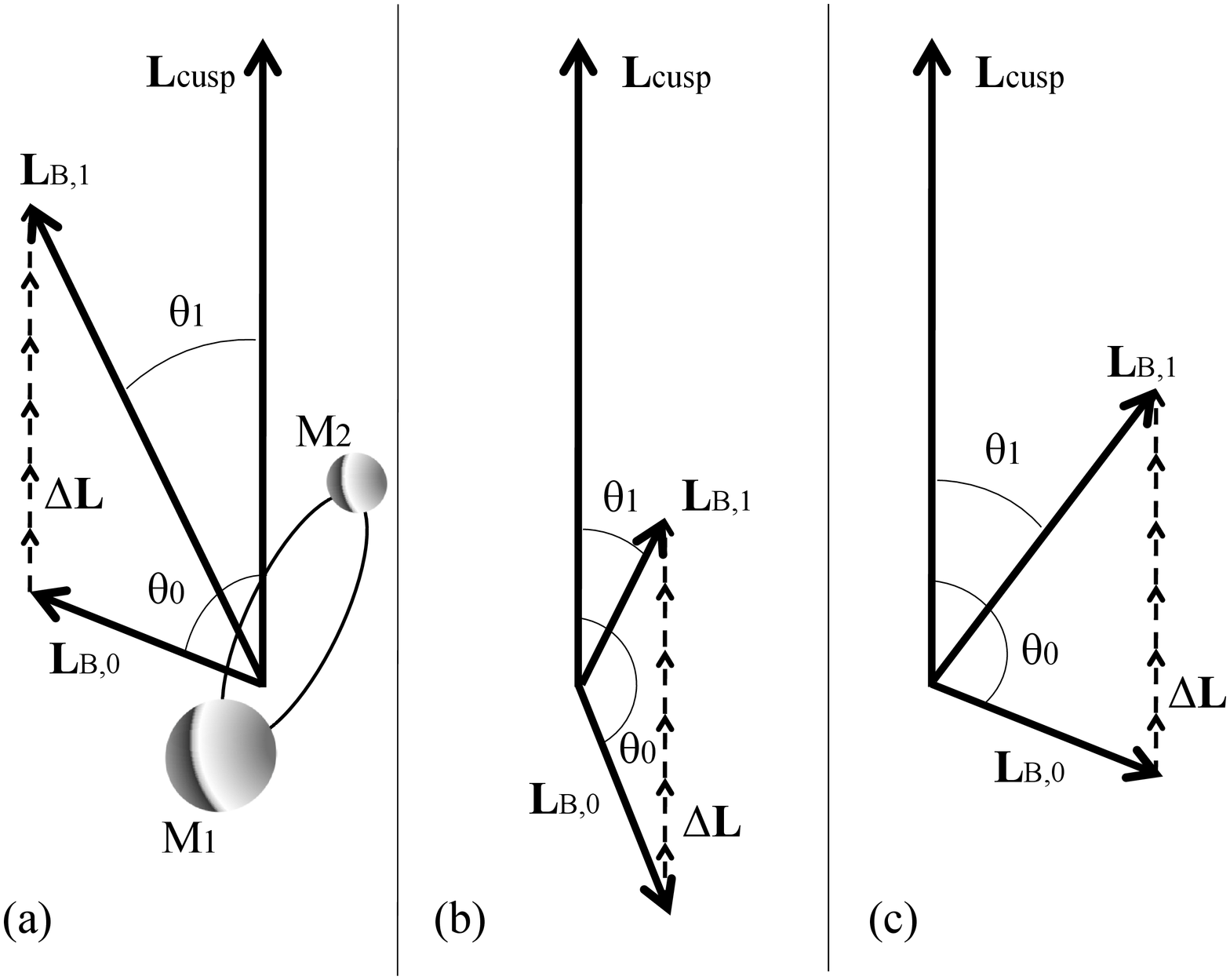}
  \end{center}
  \caption{Sketch of the model explaining the re-alignment of the
    binary plane. $\bmath{L}_{\rm B,0}$ is the initial angular
    momentum of the binary (perpendicular to the orbital plane, shown
    for clarity in the left panel). $\bmath{L}_{\rm B,1}$ is the
    angular momentum of the binary after re-alignment. $\Delta
    \bmath{L}$ is the amount of angular momentum transferred at any
    interaction with a single star (dashed line). Note that we assume
    $\Delta \bmath{L}$ parallel to the total angular momentum
    $\bmath{L}_{\rm cusp}$ of the cusp. $\theta_0$ and $\theta_1$
    indicate, respectively, the initial and final angles between
    $\bmath{L}_{\rm B}$ and $\bmath{L}_{\rm cusp}$. Panel (a) shows a
    case with $\bmath{L}_{\rm B,0}$ partially aligned with
    $\bmath{L}_{\rm cusp}$ ($\theta_0 < 90^{\circ}$). Panels (b) and
    (c) refer to the opposite case ($\theta_0 > 90^{\circ}$).  Vectors
    are not on a physical scale.
  }
  \label{fig:model}
\end{figure}
The fast alignment that we observe in the $N$-body simulations is a
direct consequence of such preferential direction of $\Delta
\bmath{l}_B$. Fig.~\ref{fig:model} shows a schematic view of our model, for
different initial orientations of the binary.  

 The evolution of $\theta$ predicted by our model is shown with dotted
 lines in Fig.~\ref{fig:theta}. Naturally, the case with $\bmath{l}_B$
 aligned with $\bmath{l}_{\rm cusp}$ does not show any
 evolution. Despite the simplifying assumptions made, the agreement
 between the initial evolution of $\theta$ in the runs and in the
 model is very good.

The predictions of the model and the results of the $N$-body runs
disagree at late times. In the simulations the binary stops
re-orienting well before perfect alignment is reached, while the model
predicts $\theta \rightarrow 0$ asymptotically. This difference can be
explained simply in terms of a depletion of stars in the centre of the
cusp. As in many previous studies \citep[e.g.][]{mm01, mm02, m06,
  mms07, gm11}, we observe the formation of a central core due to
slingshot ejections of stars.  After a core has been excavated in the
centre, the rate of encounters drastically drops and the evolution of
the BHB's orbital plane halts. This translates into a maximum angular
momentum $l_{\rm max}$ that can be transferred from the stars to the
binary . Since in our model we assumed the transferred angular
momentum to depend on the properties of the cusp and not on those of
the binary (the stars are ejected isotropically), $l_{\rm max}$ must
be the same in every run.  We computed the value of $l_{\rm max}$ that
needs to be transferred to the binary to match exactly the evolution
of the $\theta_0=120^{\circ}$ run, and we checked if such a limiting
angular momentum can account for the other runs' behaviour. The
results of this test are shown in Fig.~\ref{fig:theta}, with dashed
lines. The agreement with the $N$-body results is greatly improved by
the introduction of a maximum angular momentum that can be transferred
to the binary.

As a final note, we emphasize that the re-alignment in rotating cusps
does not depend on the $m_*/M_{\rm BHB}$ ratio, as opposed to the
non-rotating case. Since the rate of encounters {\it under every
  assumption} depends on the number density of stars, and the amount
of angular momentum transferred per encounter is proportional to
$m_*$, ${\rm d} l_{\rm B}/{\rm d} t$ depends on the mass density
$\rho$ of the cusp, but does not show any other dependence on
$m_*/M_{\rm BHB}$. This is a consequence of the fact that the
alignment in each encounter has a preferential direction (i.e. towards
the angular momentum of the cusp). The upper panel in
Fig.~\ref{fig:Ndep} shows the alignment $\Delta \theta$ experienced
over an hardening time by binaries in rotating cusps, as a function of
the number of stars in the cusp.  Different lines are for different
initial angles between $\bmath{l}_{\rm B}$ and $\bmath{l}_{\rm cusp}$.
While in the non-rotating case the ratio between the re-alignments in
the lowest and highest mass resolution runs is $\sim 30$,
in this case the re-alignment shows only statistical
fluctuations, with no systematic trends.

\section{Conclusions}
\label{sec:concl}

We studied the evolution of the orientation of the orbital plane of
unequal mass BHBs in isotropic and rotating stellar systems.  The
evolution in isotropic systems is characterized by a random walk on
long time-scales, as predicted by \citet{m02}.  The evolution in
systems in which a fraction of stars is counter-rotating with respect
to the binary was briefly discussed in \citet{sgd11}.  If the fraction
of counter-rotating stars is small, the change in the orbital plane is
consistent with the change expected for isotropic cusps. If, on the
other hand, a large fraction of stars is counter-rotating, the plane
of the binary evolves considerably and the binary angular momentum
even reverses in the case of a fully counter-rotating cusp.

Here, we investigated for the first time the evolution of the binary
plane in rotating systems in which the binary angular momentum is
initially misaligned with respect to that of the stellar cusp.  We
find that mis-aligned binaries in rotating cusps tend to align their
angular momentum with that of the cusp on a time-scale of a few
hardening times. The alignment process is linear with time, hence,
unlike the random walk observed in fully isotropic systems, does not
depend on the mass ratio between the MBHs and the stars.

The realignment of the binary plane that we observe in the simulations
may have significant implications. The direction of the spin axis of
the single MBH that results from the merger of the holes due to
emission of gravitational waves is affected by the orientation of the
binary plane before coalescence.  The spin axis, in turn, determines
the orientation of the accretion disk around the remnant black hole
via the Bardeen-Peterson effect \citep{bp75} and, in radio-loud AGNs,
the direction of the radio jet.

After the coalescence of a BHB, the spin of the remnant
$\bmath{j}_{\rm r}$ is not necessarily aligned with the spins of the
two progenitors, $\bmath{j}_1$ and $\bmath{j}_2$. If one of the two
MBHs in the binary is radio--loud (and the remnant stays radio--loud),
the change in spin direction would result in a reorientation of the
relativistic jet\footnote{This is valid in gas poor mergers. In gas
  rich environments, gas accretion causes the spins of the MBHs to
  align with the angular momentum of the binary \citep{bogdanovic07,
    dotti10, volonteri10, kesden10}, preventing any drastic change in
  the spins direction at coalescence.} \citep{me02}. The orientation
of the orbital plane of the binary has implications for the direction
of $\bmath{j}_{\rm r}$. The interaction with a rotating cusp tends to
force the binary to co-rotate with the central cusp. At coalescence,
the binary orbital angular momentum contribution to $\bmath{j}_{\rm
  r}$ will result in a spin (and possibly a jet) preferentially
aligned with the cusp angular momentum.

We do not expect full alignment, however, because: (i) the alignment
we find in the simulations is not complete, and could halt in the
absence of further refilling of stars or in the presence of isotropic
refilling, leaving misalignment angles of up to $60^{\circ}$, (ii) in
a very unequal-mass binary (e.g. $q=1/81$, like the one we study),
$\bmath{j}_{\rm r}$ depends strongly on $\bmath{j}_1$. Using the
results of \citet{rezzolla08}, if $j_1 \approx 1$, the angle between
$\bmath{j}_1$ and $\bmath{j}_{\rm r}$ is of only a few degrees, while
$j_1 \approx 0.1$ results in an angle of $\gap 20^{\circ}$.

Such a preferential alignment is in fact observed. In weak radio-loud
AGNs, \citet{brown10} found a tendency for the axis of the radio
emission to align with the minor axis of the starlight of the host, an
oblate rotationally supported elliptical. By contrast, they found no
preferred radio-optical alignment among the radio-louder objects,
\citep[see also][]{ss2009}, possibly hosted in triaxial non-rotating
ellipticals \citep[e.g.][]{kormendy09}.

\section*{Acknowledgments}

We thank Pau Amaro-Seoane and David Merritt for interesting
discussions and the anonymous referee for useful comments on the
manuscript.

\bibliographystyle{mn2e}
\bibliography{biblio}

\begin{thebibliography}{}

\bibitem[\protect\citeauthoryear{{Amaro-Seoane}, {Eichhorn}, {Porter} \&
  {Spurzem}}{{Amaro-Seoane} et~al.}{2010}]{pau2010}
{Amaro-Seoane} P.,  {Eichhorn} C.,  {Porter} E.~K.,    {Spurzem} R.,  2010,
  \mnras, 401, 2268

\bibitem[\protect\citeauthoryear{{Armitage} \& {Natarajan}}{{Armitage} \&
  {Natarajan}}{2002}]{armitage02}
{Armitage} P.~J.,  {Natarajan} P.,  2002, \apjl, 567, L9

\bibitem[\protect\citeauthoryear{{Armitage} \& {Natarajan}}{{Armitage} \&
  {Natarajan}}{2005}]{armitage05}
{Armitage} P.~J.,  {Natarajan} P.,  2005, \apj, 634, 921

\bibitem[\protect\citeauthoryear{{Bardeen} \& {Petterson}}{{Bardeen} \&
  {Petterson}}{1975}]{bp75}
{Bardeen} J.~M.,  {Petterson} J.~A.,  1975, \apjl, 195, L65+

\bibitem[\protect\citeauthoryear{{Baumgardt}, {Gualandris} \& {Portegies
  Zwart}}{{Baumgardt} et~al.}{2006}]{baum06}
{Baumgardt} H.,  {Gualandris} A.,    {Portegies Zwart} S.,  2006, \mnras, 372,
  174

\bibitem[\protect\citeauthoryear{{Begelman}, {Blandford} \& {Rees}}{{Begelman}
  et~al.}{1980}]{BBR80}
{Begelman} M.~C.,  {Blandford} R.~D.,    {Rees} M.~J.,  1980, \nat, 287, 307

\bibitem[\protect\citeauthoryear{{Berczik}, {Merritt} \& {Spurzem}}{{Berczik}
  et~al.}{2005}]{ber05}
{Berczik} P.,  {Merritt} D.,    {Spurzem} R.,  2005, \apj, 633, 680

\bibitem[\protect\citeauthoryear{{Berczik}, {Merritt}, {Spurzem} \&
  {Bischof}}{{Berczik} et~al.}{2006}]{ber06}
{Berczik} P.,  {Merritt} D.,  {Spurzem} R.,    {Bischof} H.-P.,  2006, \apjl,
  642, L21

\bibitem[\protect\citeauthoryear{{Berentzen}, {Preto}, {Berczik}, {Merritt} \&
  {Spurzem}}{{Berentzen} et~al.}{2009}]{ber09}
{Berentzen} I.,  {Preto} M.,  {Berczik} P.,  {Merritt} D.,    {Spurzem} R.,
  2009, \apj, 695, 455

\bibitem[\protect\citeauthoryear{{Bogdanovi{\'c}}, {Reynolds} \&
  {Miller}}{{Bogdanovi{\'c}} et~al.}{2007}]{bogdanovic07}
{Bogdanovi{\'c}} T.,  {Reynolds} C.~S.,    {Miller} M.~C.,  2007, \apjl, 661,
  L147

\bibitem[\protect\citeauthoryear{{Browne} \& {Battye}}{{Browne} \&
  {Battye}}{2010}]{brown10}
{Browne} I.~W.~A.,  {Battye} R.~A.,  2010, in {L.~Maraschi, G.~Ghisellini,
  R.~Della Ceca, \& F.~Tavecchio} ed., Accretion and Ejection in AGN: a Global
  View Vol.~427 of Astronomical Society of the Pacific Conference Series, {A
  Dichotomy in Radio Jet Orientations in Elliptical Galaxies}.
pp 365--+

\bibitem[\protect\citeauthoryear{{Cuadra}, {Armitage}, {Alexander} \&
  {Begelman}}{{Cuadra} et~al.}{2009}]{cuadra09}
{Cuadra} J.,  {Armitage} P.~J.,  {Alexander} R.~D.,    {Begelman} M.~C.,  2009,
  \mnras, 393, 1423

\bibitem[\protect\citeauthoryear{{Dotti}, {Colpi} \& {Haardt}}{{Dotti}
  et~al.}{2006}]{dotti06}
{Dotti} M.,  {Colpi} M.,    {Haardt} F.,  2006, \mnras, 367, 103

\bibitem[\protect\citeauthoryear{{Dotti}, {Colpi}, {Haardt} \& {Mayer}}{{Dotti}
  et~al.}{2007}]{dotti07}
{Dotti} M.,  {Colpi} M.,  {Haardt} F.,    {Mayer} L.,  2007, \mnras, 379, 956

\bibitem[\protect\citeauthoryear{{Dotti}, {Ruszkowski}, {Paredi}, {Colpi},
  {Volonteri} \& {Haardt}}{{Dotti} et~al.}{2009}]{dotti09}
{Dotti} M.,  {Ruszkowski} M.,  {Paredi} L.,  {Colpi} M.,  {Volonteri} M.,
  {Haardt} F.,  2009, \mnras, 396, 1640

\bibitem[\protect\citeauthoryear{{Dotti}, {Volonteri}, {Perego}, {Colpi},
  {Ruszkowski} \& {Haardt}}{{Dotti} et~al.}{2010}]{dotti10}
{Dotti} M.,  {Volonteri} M.,  {Perego} A.,  {Colpi} M.,  {Ruszkowski} M.,
  {Haardt} F.,  2010, \mnras, 402, 682

\bibitem[\protect\citeauthoryear{{Escala}, {Larson}, {Coppi} \&
  {Mardones}}{{Escala} et~al.}{2004}]{escala04}
{Escala} A.,  {Larson} R.~B.,  {Coppi} P.~S.,    {Mardones} D.,  2004, \apj,
  607, 765

\bibitem[\protect\citeauthoryear{{Escala}, {Larson}, {Coppi} \&
  {Mardones}}{{Escala} et~al.}{2005}]{escala05}
{Escala} A.,  {Larson} R.~B.,  {Coppi} P.~S.,    {Mardones} D.,  2005, \apj,
  630, 152

\bibitem[\protect\citeauthoryear{{Gaburov}, {Harfst} \& {Portegies
  Zwart}}{{Gaburov} et~al.}{2009}]{sapporo2009}
{Gaburov} E.,  {Harfst} S.,    {Portegies Zwart} S.,  2009, \na, 14, 630

\bibitem[\protect\citeauthoryear{{Gualandris} \& {Merritt}}{{Gualandris} \&
  {Merritt}}{2007}]{gm07}
{Gualandris} A.,  {Merritt} D.,  2007, ArXiv e-prints

\bibitem[\protect\citeauthoryear{{Gualandris} \& {Merritt}}{{Gualandris} \&
  {Merritt}}{2011}]{gm11}
{Gualandris} A.,  {Merritt} D.,  2011, ArXiv e-prints

\bibitem[\protect\citeauthoryear{{Harfst}, {Gualandris}, {Merritt}, {Spurzem},
  {Portegies Zwart} \& {Berczik}}{{Harfst} et~al.}{2007}]{Harfst2007}
{Harfst} S.,  {Gualandris} A.,  {Merritt} D.,  {Spurzem} R.,  {Portegies Zwart}
  S.,    {Berczik} P.,  2007, New Astronomy, 12, 357

\bibitem[\protect\citeauthoryear{{Kesden}, {Sperhake} \& {Berti}}{{Kesden}
  et~al.}{2010}]{kesden10}
{Kesden} M.,  {Sperhake} U.,    {Berti} E.,  2010, \apj, 715, 1006

\bibitem[\protect\citeauthoryear{{Khan}, {Just} \& {Merritt}}{{Khan}
  et~al.}{2011}]{khan11}
{Khan} F.~M.,  {Just} A.,    {Merritt} D.,  2011, \apj, 732, 89

\bibitem[\protect\citeauthoryear{{Kormendy}, {Fisher}, {Cornell} \&
  {Bender}}{{Kormendy} et~al.}{2009}]{kormendy09}
{Kormendy} J.,  {Fisher} D.~B.,  {Cornell} M.~E.,    {Bender} R.,  2009, \apjs,
  182, 216

\bibitem[\protect\citeauthoryear{{Lodato}, {Nayakshin}, {King} \&
  {Pringle}}{{Lodato} et~al.}{2009}]{lodato09}
{Lodato} G.,  {Nayakshin} S.,  {King} A.~R.,    {Pringle} J.~E.,  2009, \mnras,
  398, 1392

\bibitem[\protect\citeauthoryear{{Makino} \& {Funato}}{{Makino} \&
  {Funato}}{2004}]{mf2004}
{Makino} J.,  {Funato} Y.,  2004, \apj, 602, 93

\bibitem[\protect\citeauthoryear{{Merritt}}{{Merritt}}{2002}]{m02}
{Merritt} D.,  2002, \apj, 568, 998

\bibitem[\protect\citeauthoryear{{Merritt}}{{Merritt}}{2006}]{m06}
{Merritt} D.,  2006, \apj, 648, 976

\bibitem[\protect\citeauthoryear{{Merritt} \& {Ekers}}{{Merritt} \&
  {Ekers}}{2002}]{me02}
{Merritt} D.,  {Ekers} R.~D.,  2002, Science, 297, 1310

\bibitem[\protect\citeauthoryear{{Merritt}, {Mikkola} \& {Szell}}{{Merritt}
  et~al.}{2007}]{mms07}
{Merritt} D.,  {Mikkola} S.,    {Szell} A.,  2007, \apj, 671, 53

\bibitem[\protect\citeauthoryear{{Milosavljevi{\'c}} \&
  {Merritt}}{{Milosavljevi{\'c}} \& {Merritt}}{2001}]{mm01}
{Milosavljevi{\'c}} M.,  {Merritt} D.,  2001, \apj, 563, 34

\bibitem[\protect\citeauthoryear{{Milosavljevi{\'c}} \&
  {Merritt}}{{Milosavljevi{\'c}} \& {Merritt}}{2003}]{mm03}
{Milosavljevi{\'c}} M.,  {Merritt} D.,  2003, \apj, 596, 860

\bibitem[\protect\citeauthoryear{{Milosavljevi{\'c}}, {Merritt}, {Rest} \& {van
  den Bosch}}{{Milosavljevi{\'c}} et~al.}{2002}]{mm02}
{Milosavljevi{\'c}} M.,  {Merritt} D.,  {Rest} A.,    {van den Bosch} F.~C.,
  2002, \mnras, 331, L51

\bibitem[\protect\citeauthoryear{{Preto}, {Berentzen}, {Berczik} \&
  {Spurzem}}{{Preto} et~al.}{2011}]{preto11}
{Preto} M.,  {Berentzen} I.,  {Berczik} P.,    {Spurzem} R.,  2011, \apjl, 732,
  L26+

\bibitem[\protect\citeauthoryear{{Rezzolla}, {Barausse}, {Dorband}, {Pollney},
  {Reisswig}, {Seiler} \& {Husa}}{{Rezzolla} et~al.}{2008}]{rezzolla08}
{Rezzolla} L.,  {Barausse} E.,  {Dorband} E.~N.,  {Pollney} D.,  {Reisswig} C.,
   {Seiler} J.,    {Husa} S.,  2008, \prd, 78, 044002

\bibitem[\protect\citeauthoryear{{Roedig}, {Dotti}, {Sesana}, {Cuadra} \&
  {Colpi}}{{Roedig} et~al.}{2011}]{roedig11}
{Roedig} C.,  {Dotti} M.,  {Sesana} A.,  {Cuadra} J.,    {Colpi} M.,  2011,
  \mnras, pp 979--+

\bibitem[\protect\citeauthoryear{{Saripalli} \& {Subrahmanyan}}{{Saripalli} \&
  {Subrahmanyan}}{2009}]{ss2009}
{Saripalli} L.,  {Subrahmanyan} R.,  2009, \apj, 695, 156

\bibitem[\protect\citeauthoryear{{Sesana}}{{Sesana}}{2010}]{sesa2010}
{Sesana} A.,  2010, \apj, 719, 851

\bibitem[\protect\citeauthoryear{{Sesana}, {Gualandris} \& {Dotti}}{{Sesana}
  et~al.}{2011}]{sgd11}
{Sesana} A.,  {Gualandris} A.,    {Dotti} M.,  2011, \mnras, 415, L35

\bibitem[\protect\citeauthoryear{{Sesana}, {Haardt} \& {Madau}}{{Sesana}
  et~al.}{2007}]{sesa2007}
{Sesana} A.,  {Haardt} F.,    {Madau} P.,  2007, \apj, 660, 546

\bibitem[\protect\citeauthoryear{{Volonteri}, {G{\"u}ltekin} \&
  {Dotti}}{{Volonteri} et~al.}{2010}]{volonteri10}
{Volonteri} M.,  {G{\"u}ltekin} K.,    {Dotti} M.,  2010, \mnras, 404, 2143

\end{thebibliography}

\end{document}